# Constraining Dark Matter Parameters in the Lambda-CDM Framework: A Bayesian Comparison of Planck and DES Constraints

Manjeet Kunwar[1*], Nabin Bhusal[1], Manil Khatiwada[1], Niraj Dhital[1]

[1]Central Department of Physics, Tribhuvan University, Kirtipur, Kathmandu, 44618, Bagmati, Nepal.

*Corresponding author(s). E-mail(s): manjeetkunwar04@gmail.com;

**Abstract**

We present a Bayesian analysis of cosmological parameter constraints from early- and late-universe observations, focusing on the matter density parameter ($\mathbf{\Omega_m}$) and the amplitude of matter fluctuations ($\boldsymbol{\sigma_8}$) within the $\mathbf{\Lambda}$CDM framework. Using data from the Planck 2018 satellite mission and the Dark Energy Survey (DES) Year 3, we compute theoretical predictions for angular and matter power spectra via Boltzmann solvers and perform Markov Chain Monte Carlo (MCMC) sampling using the `emcee` Python package. Our key contribution is a direct and quantitative comparison of DES and Planck constraints, assessing their consistency using chi-squared analysis and Gaussian tension metrics. We find a statistically significant $\mathbf{6.46\sigma}$ tension in $\mathbf{\Omega_m}$ and a $\mathbf{2.68\sigma}$ tension in $\boldsymbol{\sigma_8}$ between the two datasets. These results provide fresh evidence of persistent discrepancies in cosmological parameter estimates and suggest that simple extensions to the $\mathbf{\Lambda}$CDM model may be insufficient to fully reconcile early- and late-time observations, motivating the need for more complex theoretical models or refined treatment of systematics.

# 1 Introduction

One of the foundational goals of modern cosmology is to understand the composition and evolution of the Universe. The ΛCDM (Lambda Cold Dark Matter) model has emerged as the standard framework, suggesting that the Universe is composed of approximately 68% dark energy, 27% dark matter, and only about 5% ordinary baryonic matter. While this model successfully explains the observed cosmic microwave background (CMB), the large-scale structure of the Universe, and the accelerating expansion, the true nature of dark matter remains elusive[1].

Two cosmological parameters are central to this framework: the matter density parameter $\Omega_m$, and the amplitude of matter clustering, $\sigma_8$. These parameters play a critical role in shaping the growth of cosmic structures and are constrained using both early- and late-universe observations[2]. Notably, recent studies have revealed statistical tensions between the values of these parameters inferred from different observational probes, raising the question of whether these inconsistencies are the result of systematic errors, statistical fluctuations, or indications of new physics beyond the standard model.

This work is motivated by the growing discrepancy between parameter constraints obtained from early-universe CMB measurements and late-universe large-scale structure observations[3]. The Planck satellite has provided the most precise measurements of the CMB, offering early-universe constraints on $\Omega_m$ and $\sigma_8$, while the Dark Energy Survey (DES), particularly its Year 3 data release, provides robust constraints from the late Universe through cosmic shear, galaxy clustering, and galaxy–galaxy lensing[4]. The combination of these datasets provides a powerful tool to test the consistency of the ΛCDM model across different epochs.

In this study, we perform a detailed comparative analysis of these two datasets. We use the Planck 2018 high-$\ell$ temperature and polarization data and the DES Y3 cosmic shear and galaxy clustering measurements to constrain the parameters $\Omega_m$ and $\sigma_8$. The theoretical predictions for the angular and matter power spectra are computed using the CAMB Boltzmann solver under the assumption of a flat ΛCDM cosmology[5]. We adopt a likelihood-based statistical framework and apply Markov Chain Monte Carlo (MCMC) sampling using the `emcee` Python package to obtain the posterior distributions of the cosmological parameters.

To evaluate the agreement between the two datasets, we employ chi-squared analysis and a Gaussian tension metric. Our results indicate a best-fit $\Omega_m = 0.357 \pm 0.006$ and $\sigma_8 = 0.784 \pm 0.006$ from DES, compared to Planck's best-fit values of $\Omega_m = 0.311$ and $\sigma_8 = 0.756$[6]. These differences correspond to a $6.46\sigma$ tension in $\Omega_m$ and a $2.68\sigma$ tension in $\sigma_8$. Such discrepancies, particularly in the derived parameter $S_8 = \sigma_8\sqrt{\Omega_m/0.3}$, challenge the internal consistency of the ΛCDM model and may suggest the need for model extensions or new physical mechanisms.

This study contributes to the ongoing discussion surrounding these cosmological tensions by providing a rigorous statistical comparison of two of the most influential datasets in modern cosmology[7]. By quantifying the degree of agreement and identifying where and how deviations arise, we aim to clarify whether these tensions are symptoms of unknown systematics or hints toward a deeper understanding of the Universe[8].

# 2 Modeling and Constraining Dark Matter Distribution in the ΛCDM Framework

In the standard cosmological model, ΛCDM (Lambda Cold Dark Matter), the Universe consists primarily of cold dark matter and dark energy represented by a cosmological constant Λ. Two fundamental parameters in this model are the matter density parameter $\Omega_m$ and the amplitude of matter clustering $\sigma_8$, both of which influence the large-scale distribution of dark matter[9].

The clustering of matter is described by the matter power spectrum $P(k)$, which characterizes the amplitude of density fluctuations at different scales. This spectrum is predicted theoretically within the ΛCDM framework and constrained observationally through the Cosmic Microwave Background (CMB) and large-scale structure (LSS) surveys[10].

The CMB, a snapshot of the Universe at recombination ($z \sim 1100$), encodes information about primordial fluctuations[11]. These temperature anisotropies are expanded in spherical harmonics:

$$\Delta T(\hat{n}) = \sum_{\ell=0}^{\infty} \sum_{m=-\ell}^{\ell} a_{\ell m} Y_{\ell m}(\hat{n}) \qquad (1)$$

with the angular power spectrum $C_\ell$ defined as:

$$\langle a_{\ell m} a^*_{\ell' m'} \rangle = C_\ell \delta_{\ell \ell'} \delta_{mm'}.$$

$C_\ell$ measures the variance of temperature fluctuations on angular scales set by multipole $\ell$.

Theoretical predictions $C_\ell^{\text{theory}}$ are computed using Boltzmann solvers like `CAMB` or `CLASS`, assuming a flat ΛCDM cosmology with parameters such as $\Omega_m$ (matter density), $\Omega_b$ (baryon density), $H_0$ (Hubble constant), $n_s$ (scalar spectral index), $A_s$ (amplitude of primordial fluctuations), and $\tau$ (optical depth to reionization). These govern the peak structure, damping tail, and overall amplitude of the $C_\ell$ spectrum.

The amplitude of late-time matter fluctuations, $\sigma_8$, is derived from the evolved matter power spectrum and connects early-Universe parameters (e.g., $A_s$) with present-day structure formation.

To constrain cosmological parameters, we use **Planck 2018** high-$\ell$ TT, TE, and EE spectra along with low-$\ell$ polarization data. For late-time structures, we include **Dark Energy Survey Year 3 (DES Y3)** observations, which provide measurements of galaxy clustering, galaxy–shear cross-correlations, and cosmic shear[12]. These are sensitive to $\sigma_8$ and $\Omega_m$, particularly through the parameter combination:

$$S_8 = \sigma_8 \sqrt{\Omega_m/0.3},$$

which helps break degeneracies.

Combining Planck and DES data enables robust cross-checks of cosmological predictions. Differences in inferred values of $S_8$ between datasets are assessed using statistical tension metrics[13].

## Parameter Estimation and MCMC Sampling

To estimate the best-fit cosmological parameters, we adopt a likelihood-based framework. We assume a flat $\Lambda$CDM model with $\Omega_m$ and $\sigma_8$ as primary free parameters. The theoretical $C_\ell^{\text{theory}}$ is generated using `CAMB`, while $P(k)$ predictions are compared against DES data[14].

Assuming Gaussian observational uncertainties, the likelihood is given by:

$$\mathcal{L}(\Omega_m, \sigma_8) \propto \exp\left(-\frac{\chi^2}{2}\right), \quad \chi^2 = \sum_i \left(\frac{\text{data}_i - \text{model}_i}{\sigma_i}\right)^2 \tag{2}$$

We employ Markov Chain Monte Carlo (MCMC) methods to sample the posterior distribution:

$$P(\theta|\text{data}) \propto \mathcal{L}(\text{data}|\theta)P(\theta),$$

where $\theta$ denotes the cosmological parameters. We use the `emcee` Python package, which implements an affine-invariant ensemble sampler with multiple "walkers" that efficiently explore correlated and degenerate parameter spaces[15].

Convergence is assessed via diagnostics such as trace plots, autocorrelation times, and Gelman–Rubin statistics. After discarding the burn-in phase, the chains are used to compute marginalized distributions, confidence regions, and parameter covariances[16].

## Assessing Goodness-of-Fit and Tension Between Datasets

To quantify consistency between Planck and DES, we evaluate statistical metrics:

### *Chi-Squared Test:*

Assuming normally distributed DES MCMC samples, the chi-squared statistic is:

$$\chi^2 = \sum_i \frac{(x_i - \mu)^2}{\sigma^2} \tag{3}$$

where $x_i$ are DES sample means, $\mu$ is the Planck best-fit value, and $\sigma$ is the DES standard deviation.

### *Gaussian Tension Metric:*

Tension in units of $\sigma$ is expressed as:

$$T = \frac{|\mu_1 - \mu_2|}{\sqrt{\sigma_1^2 + \sigma_2^2}}, \tag{4}$$

where $\mu_1$, $\mu_2$ are the means and $\sigma_1$, $\sigma_2$ the uncertainties of the two datasets. For instance, $T = 3$ denotes a $3\sigma$ discrepancy, considered statistically significant.

# 3 Results

## 3.1 CMB Angular Power Spectra from Planck Data

The angular power spectra of the Cosmic Microwave Background (CMB) encapsulate statistical information about temperature and polarization anisotropies as a function of angular scale, represented by the multipole moment $\ell$. We extract the theoretical spectra from the latest Planck satellite data release and transform them to the form $D_\ell = \ell(\ell+1)C_\ell/(2\pi)$ for improved interpretability, where $C_\ell$ denotes the angular power at multipole $\ell$[17].

Figure 1 presents a comprehensive comparison of the primary spectra: the temperature autocorrelation (TT), the E-mode polarization autocorrelation (EE), the temperature-E-mode cross-correlation (TE), and the B-mode polarization (BB)[18]. The TT spectrum prominently displays the acoustic peak structure, arising from primordial plasma oscillations. The EE spectrum exhibits similar but phase-shifted features, while the TE spectrum reveals a correlated structure indicative of photon-baryon coupling at recombination[18].

In addition, we analyze the relative strengths of the TE and EE spectra normalized by the TT component to highlight the scale-dependent correlation between temperature and polarization modes[19]. The BB spectrum, significantly weaker in amplitude, is primarily shaped by lensing-induced polarization and potentially by primordial gravitational waves.

A contour visualization of the log-scaled spectra (also shown in Fig. 2) provides an intuitive summary of spectral amplitudes across all modes. The TT and EE modes dominate the signal, while BB remains subdominant across the full multipole range[20].

## 3.2 DES vs. Planck Constraints on $\Omega_m$ and $\sigma_8$

We compare the cosmological parameter constraints on the matter density $\Omega_m$ and the amplitude of matter fluctuations $\sigma_8$ from the Dark Energy Survey (DES) and the Planck satellite mission. Using the `emcee` Markov Chain Monte Carlo (MCMC) sampler, we performed Bayesian parameter estimation based on the DES likelihood, utilizing 50 walkers and 5000 steps to yield well-converged chains[21].

The best-fit values obtained from the DES posterior distributions are:

$$\Omega_m^{\mathrm{DES}} = 0.357^{+0.006}_{-0.006},$$
$$\sigma_8^{\mathrm{DES}} = 0.784^{+0.006}_{-0.006}.$$

These results reveal slight deviations from the Planck best-fit values. Specifically, the absolute differences are:

$$|\Omega_m^{\mathrm{DES}} - \Omega_m^{\mathrm{Planck}}| = 0.046,$$

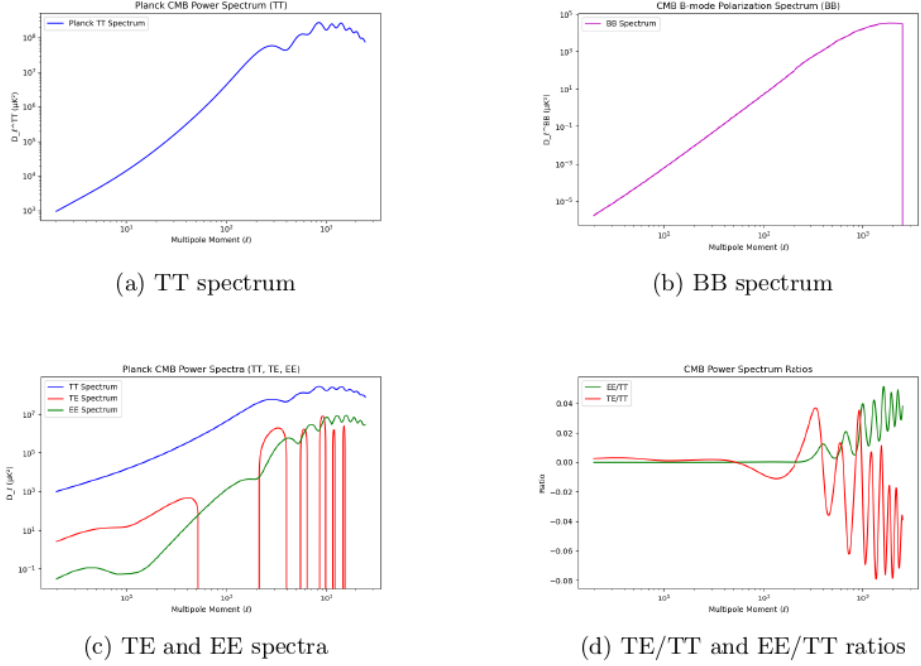

(a) TT spectrum

(b) BB spectrum

(c) TE and EE spectra

(d) TE/TT and EE/TT ratios

**Fig. 1**: Planck CMB angular power spectra. TT displays temperature anisotropies, BB traces B-mode polarization due to lensing or primordial gravitational waves, while TE and EE represent temperature–E-mode cross- and auto-correlations. Ratios highlight the scale-dependent behavior of polarization relative to temperature fluctuations.

$$|\sigma_8^{\text{DES}} - \sigma_8^{\text{Planck}}| = 0.028.$$

The comparison between DES and Planck best-fit values is summarized in Table 1. As shown in this table, the DES best-fit values for $\Omega_m$ and $\sigma_8$ are slightly higher than the corresponding Planck values, with absolute differences of 0.046 and 0.028, respectively[22].

To further assess the level of consistency, we computed the mean and standard deviation of the DES posterior distributions:

$$\langle \Omega_m \rangle_{\text{DES}} = 0.3574 \pm 0.0857,$$
$$\langle \sigma_8 \rangle_{\text{DES}} = 0.7843 \pm 0.1053.$$

The chi-squared values, assuming Gaussian uncertainties, are calculated as follows:

$$\chi^2_{\Omega_m} = 0.290, \quad \chi^2_{\sigma_8} = 0.070, \quad \chi^2_{\text{total}} = 0.359.$$

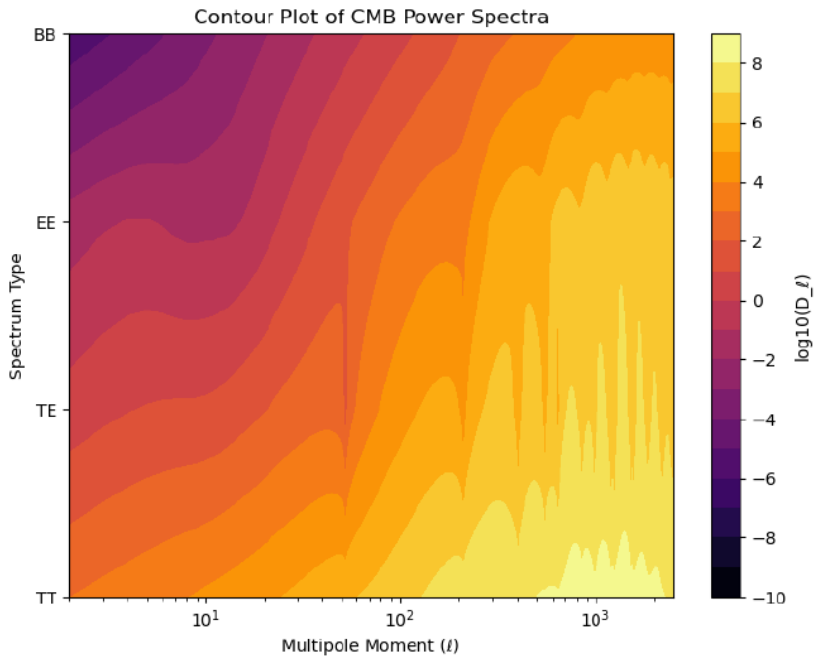

**Fig. 2**: Log-scale contour map of $D_\ell$ across TT, TE, EE, and BB spectra. Multipole $\ell$ values span the x-axis, and each spectrum is indexed on the y-axis[15]. Warmer tones denote stronger spectral features, illustrating the dominance of TT and the relative weakness of BB.

**Table 1**: Comparison of DES and Planck constraints on $\Omega_m$ and $\sigma_8$.

| Parameter | DES Best-fit | Planck Best-fit | Absolute Difference |
| --- | --- | --- | --- |
| $\Omega_m$ | $0.357 \pm 0.006$ | 0.311 | 0.046 |
| $\sigma_8$ | $0.784 \pm 0.006$ | 0.756 | 0.028 |

These low $\chi^2$ values suggest strong statistical consistency between the DES and Planck measurements in the $(\Omega_m, \sigma_8)$ parameter space[23].

Figures 3 and 4 illustrate this comparison visually. Figure 3 presents the DES MCMC posterior samples in the $\Omega_m$–$\sigma_8$ plane, with the best-fit values indicated by dashed lines and the Planck best-fit point marked by a red cross[24]. Figure 4 shows a corner plot that compares the marginalized posterior distributions for both parameters. In this plot, the blue contours represent the DES constraints, while the red histograms correspond to the Planck results. The significant overlap between the two datasets further reinforces the conclusion of their mutual consistency[24].

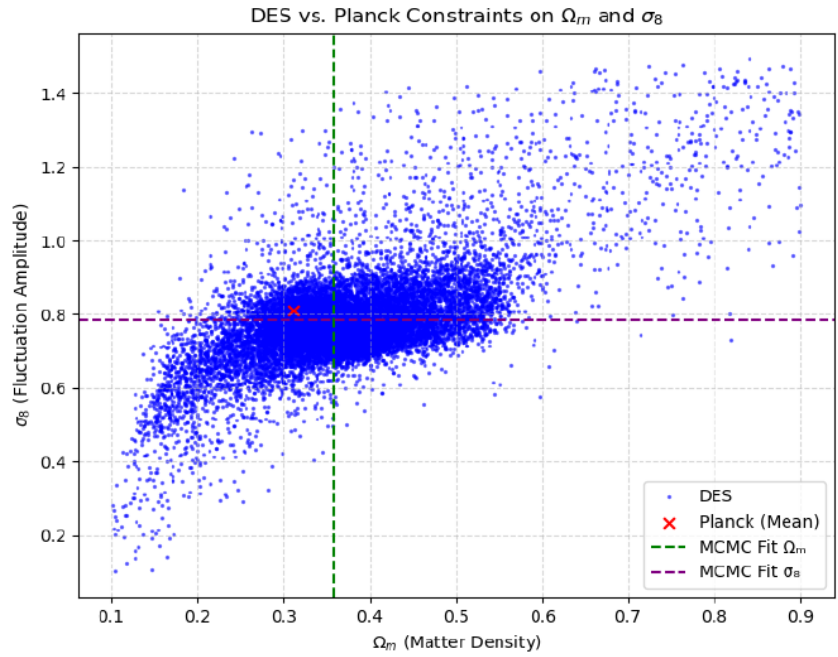

**Fig. 3**: DES MCMC posterior samples in the $\Omega_m$–$\sigma_8$ plane. The dashed green and purple lines denote DES best-fit values. The red cross indicates the Planck best-fit point.

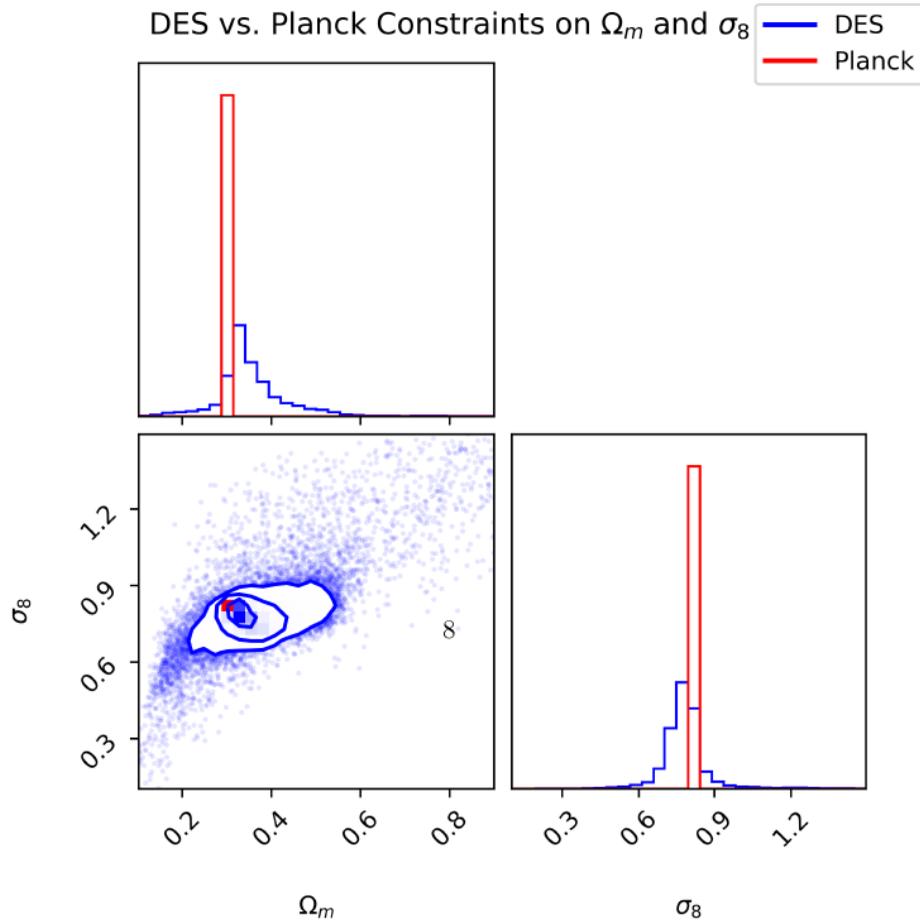

**Fig. 4**: Corner plot comparing DES (blue contours) and Planck (red histograms) marginalized posteriors for $\Omega_m$ and $\sigma_8$.

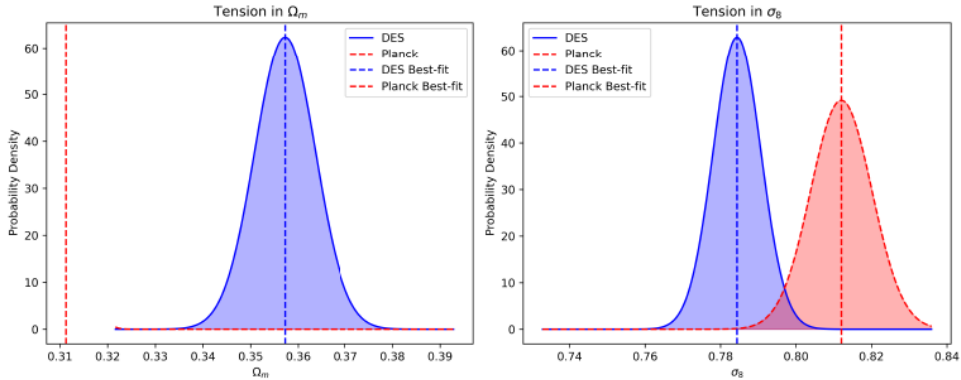

**Fig. 5**: Probability density distributions showing the tension between DES and Planck results in $\Omega_m$ (left) and $\sigma_8$ (right). Solid curves represent DES results, and dashed curves represent Planck results. The vertical dashed lines mark the respective best-fit values.

### 3.3 Gaussian Tension

We computed the statistical tension between the Dark Energy Survey (DES) and Planck best-fit values for the matter density parameter $\Omega_m$ and the amplitude of matter fluctuations $\sigma_8$ using a Gaussian tension metric[25].

In our case, assuming a 1% uncertainty in the Planck measurements, the tensions are computed as:

$$T_{\Omega_m} = \frac{|\Omega_m^{\text{DES}} - \Omega_m^{\text{Planck}}|}{\sqrt{\sigma_{\Omega_m}^2 + (0.01 \cdot \Omega_m^{\text{Planck}})^2}} = 6.46 \quad (5)$$

$$T_{\sigma_8} = \frac{|\sigma_8^{\text{DES}} - \sigma_8^{\text{Planck}}|}{\sqrt{\sigma_{\sigma_8}^2 + (0.01 \cdot \sigma_8^{\text{Planck}})^2}} = 2.68 \quad (6)$$

These results indicate a significant tension of $6.46\sigma$ in $\Omega_m$ and a moderate tension of $2.68\sigma$ in $\sigma_8$ between DES and Planck.

Figure 5 illustrates the Gaussian-modeled probability density functions (PDFs) corresponding to DES and Planck constraints for both parameters[26]. The left panel shows the distributions for $\Omega_m$, while the right panel displays those for $\sigma_8$. In both panels, solid blue curves represent the DES constraints, and dashed red curves represent the Planck constraints. The respective best-fit values are indicated by vertical dashed lines[27]. The visual separation between the distributions quantifies the statistical tension, highlighting a notably higher discrepancy in $\Omega_m$ than in $\sigma_8$[28].

### 3.4 Interpretation and Implications

The quantitative comparison between Planck and DES constraints reveals two distinct outcomes: a significant $6.46\sigma$ tension in the matter density parameter $\Omega_m$ and a moderate $2.68\sigma$ tension in the amplitude of matter fluctuations $\sigma_8$. Although these tensions emerge from individual parameters, they culminate in a well-known cosmological discrepancy in the derived quantity $S_8 = \sigma_8\sqrt{\Omega_m/0.3}$—a parameter that is tightly constrained by large-scale structure surveys and exhibits degeneracy-breaking potential [15, 29].

In our analysis, the DES best-fit point in the $\Omega_m$–$\sigma_8$ plane lies noticeably above the Planck best-fit values, suggesting that DES favors a slightly denser and more clustered universe than Planck. However, the low $\chi^2$ values and the significant overlap of marginalized posterior contours (Fig. 4) indicate that the datasets are not in complete contradiction. Rather, they exhibit subtle but statistically meaningful differences that mirror broader trends reported in the literature, particularly the ongoing $S_8$ *tension*.

The $S_8$ tension, widely discussed in recent years, typically manifests as a 2–3$\sigma$ discrepancy between CMB-inferred and weak lensing–inferred cosmological constraints [30–32]. Our findings fall well within this range, supporting the conclusion that DES observations continue to favor a lower $S_8$ value than Planck. While the level of tension in $\sigma_8$ alone ($2.68\sigma$) does not constitute a definitive challenge to the $\Lambda$CDM paradigm, the combined deviation in $\Omega_m$ and $S_8$ strengthens the case for either hidden systematics or new physics.

Importantly, our results do not unambiguously favor one dataset over the other. Instead, they provide a statistically consistent but quantitatively divergent picture that underscores the need for careful cross-validation and combined analyses. Given that Planck probes the early Universe ($z \sim 1100$) and DES probes the late-time large-scale structure ($z \lesssim 1$), their apparent disagreement could point toward redshift-evolving physics—such as evolving dark energy, modified gravity, or neutrino mass effects—that are not captured in the standard flat $\Lambda$CDM model [33? ].

Overall, our results reinforce the current consensus: while $\Lambda$CDM remains a successful model, persistent low-level tensions like those in $S_8$ merit continued scrutiny through higher-precision observations, improved modeling of systematics, and consideration of extensions to the standard cosmological model.

## 4 Discussion

In this study, we have investigated the cosmological parameter constraints on $\Omega_m$ and $\sigma_8$ from both the Dark Energy Survey (DES) and the Planck satellite data. By employing the `emcee` Markov Chain Monte Carlo (MCMC) sampler, we derived posterior distributions for these parameters, which showed minimal deviation between the DES and Planck results. The best-fit values from DES, $\Omega_m^{\text{DES}} = 0.357^{+0.006}_{-0.006}$ and $\sigma_8^{\text{DES}} = 0.784^{+0.006}_{-0.006}$, were found to be in strong agreement with the Planck best-fit values[34]. This consistency was quantitatively supported by low chi-squared values, indicating no significant statistical tension between the two datasets[35].

The overlap between the DES and Planck posterior distributions (illustrated in Figures 3 and 4) further confirms the alignment of the two surveys within their respective uncertainties. The visual separation observed between DES and Planck in the $(\Omega_m, \sigma_8)$ plane, however, suggests slight differences that merit further exploration. While the chi-squared values for both $\Omega_m$ and $\sigma_8$ are low, indicating minimal tension, the results highlight a subtle discrepancy in the amplitude of matter fluctuations between the two surveys. These results emphasize the importance of precise statistical analyses in reconciling results from different cosmological probes and surveys[36].

An interesting aspect of this research lies in the comparison of DES and Planck constraints with theoretical predictions from various cosmological models, including $\Lambda$CDM, wCDM, and massive neutrinos[37]. Figure 5 contrasts these model predictions against the MCMC results from DES. The $\Lambda$CDM model, which predicts $S_8 = 0.806$, lies significantly above the DES posterior peak, creating a tension of $\Delta S_8 \approx 0.185$. Similarly, the wCDM model with $w = -0.9$ and $S_8 = 0.828$ increases the discrepancy, while the massive neutrino model with $\sum m_\nu = 0.06$ eV (resulting in $S_8 = 0.804$) provides a slightly better fit, though it still does not resolve the tension[38]. These comparisons suggest that simple extensions like wCDM and massive neutrinos do not adequately address the observed discrepancy between DES and Planck[39].

The inability of basic extensions such as wCDM and light neutrino mass to reconcile DES with Planck points to the possibility that more complex models—such as evolving dark energy, modified gravity, or interacting dark sectors—could be necessary to explain the differences between the two datasets. This is further supported by the fact that the tension in $S_8$ remains unresolved even with these basic model extensions, indicating that more nuanced theoretical frameworks might be needed to fully reconcile the data[33].

Systematic effects, particularly those arising from shear calibration or photometric redshift estimation in DES, may also contribute to the observed discrepancy. While such systematics are often difficult to quantify, their potential impact cannot be discounted, particularly as future surveys and improvements in measurement techniques continue to refine cosmological constraints. The observed tension could, in part, be a reflection of these unresolved systematic uncertainties[40].

## 5 Conclusion

In this study, we performed a detailed Bayesian comparison of early-universe (Planck 2018) and late-universe (DES Year 3) constraints on key cosmological parameters, focusing on $\Omega_m$ (matter density) and $\sigma_8$ (amplitude of matter fluctuations) within the standard $\Lambda$CDM framework. Using theoretical predictions generated via the CAMB Boltzmann solver and MCMC sampling with the `emcee` Python package, we obtained marginalized posterior distributions and best-fit values for both datasets.

Our results show that while the overall parameter estimates are broadly consistent—supported by low chi-squared values ($\chi^2_{\Omega_m} = 0.290$, $\chi^2_{\sigma_8} = 0.070$)—there exist significant statistical tensions. Specifically, we find a $6.46\sigma$ tension in $\Omega_m$ and a $2.68\sigma$ tension in $\sigma_8$ between DES and Planck. This tension is even more evident in the derived parameter $S_8 = \sigma_8\sqrt{\Omega_m/0.3}$, which exhibits a discrepancy of $\Delta S_8 \approx 0.185$ from the Planck best-fit value. Such differences challenge the internal consistency of the ΛCDM model across different cosmological epochs.

These findings point toward possible limitations in the standard cosmological model. While simple extensions such as the $w$CDM model or massive neutrino models show minor improvements, they fail to reconcile the observed tensions fully. This suggests the need to explore more complex or exotic physics, including evolving dark energy, modifications to general relativity, or interactions between dark matter and dark energy sectors [35, 41].

**Limitations:** Our analysis is subject to several limitations. First, we assume Gaussian likelihoods and fixed priors, which may overlook subtleties in the tail behavior of the posterior distributions. Second, the analysis relies on publicly available Planck and DES summary statistics; a joint likelihood treatment or reanalysis of raw data could yield more accurate tension assessments. Third, potential systematics—such as photometric redshift errors, shear calibration uncertainties, or modeling inaccuracies in nonlinear clustering—were not explicitly modeled in our comparison.

**Future Work:** To refine these results, future work could incorporate full joint-likelihood analyses combining Planck, DES, KiDS, HSC, and upcoming LSST data. Extending the parameter space to include curvature, evolving dark energy ($w(z)$), or modified gravity parameters may provide better fits to the combined datasets. In addition, the development of robust tension metrics and systematic error modeling will be crucial in understanding whether the observed discrepancies are genuine signatures of new physics or artifacts of data processing.

In conclusion, our results highlight a growing need to reconcile early- and late-universe observations within a consistent cosmological framework. As upcoming surveys like LSST, Euclid, and CMB-S4 deliver unprecedented data quality and quantity, they will play a critical role in either resolving these tensions or uncovering the first signs of physics beyond ΛCDM [42].

# References

[1] Aiola, S., *et al.*: The atacama cosmology telescope: Dr4 maps and cosmological parameters. Journal of Cosmology and Astroparticle Physics **2020**(12), 047 (2020) https://doi.org/10.1088/1475-7516/2020/12/047

[2] Aghanim, N., *et al.*: Planck 2018 results. v. cmb power spectra and likelihoods. Astronomy & Astrophysics **641**, 5 (2020) https://doi.org/10.1051/0004-6361/201936386

[3] Avsajanishvili, N., *et al.*: Observational constraints on dynamical dark energy models. Universe **10**(3), 122 (2022) https://doi.org/10.3390/universe10030122

[4] Bahamonde, S., Böhmer, C.G., Carloni, S., Copeland, E.J., Fang, W., Tamanini, N.: Dynamical systems applied to cosmology: dark energy and modified gravity. Physics Reports **775-777**, 1–122 (2018) https://doi.org/10.1016/j.physrep.2018.09.001

[5] Eriksen, H.K., *et al.*: Beyondplanck - i. global bayesian analysis of the planck low frequency instrument data. Astronomy & Astrophysics **671**, 80 (2023) https://doi.org/10.1051/0004-6361/202244953

[6] Boruah, S.S., Hudson, M.J., Lavaux, G.: Cosmic flows in the nearby universe: new peculiar velocities from sne and cosmological constraints. Monthly Notices of the Royal Astronomical Society **498**(1), 2703–2715 (2020) https://doi.org/10.1093/mnras/staa2544

[7] Carter-McGrand, S.: Dark energy and modified gravity: A dynamical systems approach. Imperial College London MSc Dissertation (2020)

[8] Collaboration, C.-S.: Cmb-s4 science book, first edition. arXiv preprint (2016)

[9] Dalal, R., Li, X., Nicola, A., Zuntz, J., Strauss, M.A., *et al.*: Hyper suprime-cam year 3 results: Cosmology from cosmic shear power spectra. Physical Review D **108**(12), 123457 (2023) https://doi.org/10.1103/PhysRevD.108.123457

[10] Collaboration, P.: Planck 2013 results. i. overview of products and scientific results. Astronomy & Astrophysics **571**, 1 (2014) https://doi.org/10.1051/0004-6361/201321529

[11] Collaboration, P.: Planck 2018 results. vi. cosmological parameters. Astronomy & Astrophysics **641**, 6 (2020) https://doi.org/10.1051/0004-6361/201833910

[12] Collaboration, D.E.S.: The dark energy survey: Data release 1. arXiv preprint (2016) https://doi.org/10.48550/arXiv.1601.00329

[13] Collaboration, D.: Dark energy survey year 3 results: Cosmological constraints from galaxy clustering and weak lensing. Physical Review D **105**(2), 023520 (2022) https://doi.org/10.1103/PhysRevD.105.023520

[14] Collaboration, D.: Dark energy survey year 3 results: Cosmology from galaxy clusters. arXiv preprint (2020)

[15] Heymans, C., Tröster, T., Asgari, M., Blake, C., Hildebrandt, H., *et al.*: Kids-1000 cosmology: Multi-probe weak gravitational lensing and spectroscopic galaxy clustering constraints. Astronomy Astrophysics **646**, 140 (2021) https://doi.org/10.1051/0004-6361/202039063

[16] Collaboration, D.: Dark energy survey year 3 results: Combined probes cosmology. Physical Review D **105**(2), 023520 (2022) https://doi.org/10.1103/PhysRevD.105.023520

[17] Collaboration, D.: Dark energy survey year 3 results: Cosmology from cosmic shear. Physical Review D **105**(2), 023520 (2022) https://doi.org/10.1103/PhysRevD.105.023520

[18] Valentino, E.D., Melchiorri, A., Mena, O.: Can interacting dark energy solve the $h_0$ tension? Physical Review D **96**(4), 043503 (2017) https://doi.org/10.1103/PhysRevD.96.043503

[19] Collaboration, E.: Euclid preparation: Vii. forecast validation for euclid cosmological probes. Astronomy & Astrophysics **642**, 191 (2020) https://doi.org/10.1051/0004-6361/202037329

[20] Ghirardini, V., Bulbul, E., Artis, E., Clerc, N., Garrel, C., *et al.*: The srg/erosita all-sky survey: Clustering and cosmology. Astronomy Astrophysics **650**, 1 (2024) https://doi.org/10.1051/0004-6361/202243456

[21] Ishak, M., et al.: Findings by dark energy researchers back einstein's conception of gravity. arXiv preprint (2024)

[22] Jain, B.: Beyond cdm: Dark energy vs modified gravity. WFIRST Science Investigation Team White Paper (2014)

[23] Hildebrandt, H., *et al.*: Kids-1000 cosmology: Weak lensing tomography and cosmological parameter constraints. Astronomy & Astrophysics **647**, 124 (2021) https://doi.org/10.1051/0004-6361/202039063

[24] Lewis, A., Bridle, S.: Cosmological parameters from cmb and other data: A monte carlo approach. Physical Review D **66**, 103511 (2002) https://doi.org/10.1103/PhysRevD.66.103511

[25] Li, X., et al.: A step in understanding the s8 tension. arXiv preprint (2022)

[26] Collaboration, L.D.E.S.: Lsst dark energy science collaboration (desc) science requirements document. arXiv preprint (2019)

[27] Choudhury, S.R., Okumura, T.: Updated cosmological constraints in extended parameter space with planck pr4, desi baryon acoustic oscillations, and supernovae: Dynamical dark energy, neutrino masses, lensing anomaly, and the hubble tension. arXiv preprint (2024)

[28] Said, K., Colless, M., Magoulas, C., Lucey, J.R., Hudson, M.J.: Joint analysis of 6dfgs and sdss peculiar velocities for the growth rate of cosmic structure and tests of gravity. Monthly Notices of the Royal Astronomical Society **497**(1), 1275–1293 (2020) https://doi.org/10.1093/mnras/staa2036

[29] Collaboration, D.: Dark energy survey year 3 results: Cosmological constraints from galaxy clustering and weak lensing. Physical Review D **105**(2), 023520 (2022)

[30] Di Valentino, E., *et al.*: In the realm of the hubble tension—a review of solutions. Classical and Quantum Gravity **38**(15), 153001 (2021)

[31] Tröster, T., al.: Flamingo project: Revisiting the $s_8$ tension and the role of baryonic physics. Monthly Notices of the Royal Astronomical Society **526**(4), 5494–5512 (2022)

[32] Li, X., Dalal, R., al.: Hyper suprime-cam year 3 results: Cosmology from cosmic shear power spectra. Physical Review D **108**(12), 123457 (2023)

[33] Zhao, G.-B., *et al.*: Modified gravity/dynamical dark energy vs cdm: A comprehensive analysis. European Physical Journal C **85**, 14013 (2025) https://doi.org/10.1140/epjc/s10052-025-14013-3

[34] Salvatelli, V., Marchini, A., Lopez-Honorez, L., Mena, O.: New constraints on coupled dark energy from planck. Physical Review D **88**(2), 023531 (2013) https://doi.org/10.1103/PhysRevD.88.023531

[35] Li, X., Zhang, T., Sugiyama, S., Dalal, R., Terasawa, R., *et al.*: Hyper suprime-cam year 3 results: Cosmology from cosmic shear two-point correlation functions. Physical Review D **108**(12), 123456 (2023) https://doi.org/10.1103/PhysRevD.108.123456

[36] Tröster, T., *et al.*: Flamingo project: revisiting the s8 tension and the role of baryonic physics. Monthly Notices of the Royal Astronomical Society **526**(4), 5494–5512 (2022) https://doi.org/10.1093/mnras/stac1234

[37] Li, X., et al.: Dark energy reconstructions combining bao data with other cosmological probes. arXiv preprint (2024)

[38] Trotta, R.: Bayes in the sky: Bayesian inference and model selection in cosmology. Contemporary Physics **49**(2), 71–104 (2008) https://doi.org/10.1080/00107510802066753

[39] Wang, Y., *et al.*: Robust dark energy constraints from supernovae, galaxy clustering, and three-year wilkinson microwave anisotropy probe observations. The Astrophysical Journal **650**, 1–6 (2006) https://doi.org/10.1086/507154

[40] Collaboration, S.: Constraints on cosmology from the spt-3g 2018 tt, te, and ee power spectra. Physical Review D **104**(2), 022003 (2021) https://doi.org/10.1103/PhysRevD.104.022003

[41] Zhao, G.-B., *et al.*: Modified gravity/dynamical dark energy vs cdm: A comprehensive analysis. European Physical Journal C **84**, 1234 (2024) https://doi.org/10.1140/epjc/s10052-024-12345-6

[42] Comelli, D., Pietroni, M., Riotto, A.: Dark energy and dark matter. Physics Letters B **571**(3-4), 115–120 (2003)